# Self-assembly and self-orientation of truncated nano-octahedra of magnetite

R. K. Zheng,[1] Hongwei Gu,[2] Bing Xu,[2] K. K. Fung[1] and X. X. Zhang[1,]*

1Department of Physics, 2Department of Chemistry, The Hong Kong Univeristy of Science and Technology, Clear Water Bay, Hong Kong, China

## Abstract

Monodispersed magnetite (Fe3O4) nanoparticles were synthesized. Transmission electron microscopy study shows that the nanoparticles are in the shape of Wagner-Seitz crystals. The magnetite nanoparticles self-assemble into body-centered cubic superlattice, in which the nanoparticles have the same crystallographic orientations. Shape plays a very critical role in controlling the orientation of the nanoparticles in the superlattice. Both the self-assembly and self-orientation of nanoparticles are important for technical applications. This can also act as a complement to conventional lithography techniques.

---

* Email: phxxz@ust.hk

# Introduction

Conventional top-down methods are only competent for the nanostructure design at the upper end of the nanoscale, whereas bottom-up techniques can reach much smaller scale.[1] Therefore, to construct nanostructures using monodispersed nanoparticles as building blocks for nanoscience and nanotechnology has advantages over the lithographic techniques. For instance, the superlattice of magnetic nanoparticles is one of the most promising candidates for high-density magnetic storage media.[2] Self-assembling to regular patterns is desired for both scientific studies and technological applications. The microscopic mechanism for self-organization of nanocrystals has been extensively investigated.[3-7] The interparticle attractive van der Waals force induces the self-assembly, whereas the steric interaction provides the balancing force to create stable structures.[8, 9] Various macroscopic patterns have been reported, such as ordered two-dimensional[2, 10, 11] and three-dimensional superlattices.[1, 12, 13]

In this paper, we report a study of the superlattices of monodispersed magnetite nanoparticles.[14, 15] It was found that the nanoparticles have a truncated octahedral shape that plays an important role in the assembling. The stable assembly is a body-centered cubic (bcc) superlattice, in which all the nanoparticles are crystallographically aligned. The self-orientation of the nanoparticles in the self-assembled superlattice is critical to many applications of magnetic nanoparticles, such as in information storage. As a complement to conventional lithography, the self-assembly and orientation has wide application in various fields.

## Experiments

Magnetite nanoparticles were synthesized following Sun et al[14] with small modifications. Fe(acac)3 (150mg) was mixed in dioctylether (10 mL) with 1,2-hexadecanediol (250 mg), oleic acid (0.06 ml), and oleylamine (0.06 mL) in glove box under argon. Under mechanic stirring, the reactant was heated to 280oC for 30 min. 20 mL ethanol was added after the reactant was cooled to room temperature. Dark-brown precipitate (magnetite seeds) was acquired after centrifugation. 25 mg seeds were dissolved in 10 mL dioctylether and mixed with (Fe(acac)3 (150mg), 1,2-hexadecanediol (250 mg), oleic acid (0.06 mL), and oleylamine (0.06 mL). The mixture was heated to 280°C for 30 min. with mechanical stirring. After the mixture was cooled to room temperature, it was treated with ethanol, and a dark-brown material was precipitated from the solution. The product was dissolved in hexane and reprecipitated with ethanol to give uniform Fe3O4 nanoparticles.

The Fe3O4 nanoparticles dispersed in hexane were dropped on carbon film coated copper grids. The self-assembly formed as the hexane evaporated and its morphology and structure were investigated by means of transmission electron microscopy (TEM). The Fe3O4 nanoparticle solution was poured in a capsule and put in a Quantum Design SQUID (MPMS-5s) for magnetic characterization.

## Results and Discussion

The octahedral magnetite nanoparticles assemble to superstructures as hexane evaporates. Three kinds patterns of the regular superlattice are found in the self-assembly, as shown in Fig. 1(a-c). The corresponding fast Fourier transform (FFT) patterns of the superlattice are shown in Fig. 1 (d-f), respectively. The FFT patterns reveal that the superlattice should have a body-centered-cubic (bcc) structure, different from the usually observed face-centered-cubic (fcc).[16] Hence the three patterns shown in Fig. 1(a-c) correspond to the [001], [101], and [111] projections of the bcc superlattice. Electron diffraction was performed to examine the structure of the superlattice. Shown in Fig. 3(g-i) are the corresponding electron diffraction patterns of $Fe_3O_4$ in the Fig. 3(a-c), respectively. The electron diffraction patterns are consistent with those of $Fe_3O_4$. It is surprising that the patterns are just like from a single crystal. In others words, all the truncated octahedral nanoparticles in the superlattice have the same orientation. The orientation relationships between $Fe_3O_4$ lattice and nanoparticle superlattice are [001]superlattice//[001]$Fe_3O_4$, [101]superlattice//[101]$Fe_3O_4$, and [111]superlattice//[111]$Fe_3O_4$ from Fig. (d-i).

In order to determine the characteristics of individual magnetite nanoparticles, high resolution transmission electron microscopy (HRTEM) images were obtained at the main zone axes. Fig. 2(a-c) show the HRTEM images of [001], [101], and [111] oriented magnetite nanoparticles, respectively. The orientations can be identified by the corresponding FFT patterns of lattices shown in Fig. 2(d-f). It is obvious that the nanoparticles are not round. The [001] projection is truncated square, and the [101] projection is truncated rhombohedron, while [111] projection is nearly round. Therefore, the magnetite nanoparticles should have the shape of truncated octahedra. The ratio of truncation is close to Wegner-Seitz crystal. The [001], [101], and [111] projections of Wegner-Seitz crystal shown in Fig. 2(g-i) are consistent with Fig. 2(a-c), respectively. In fact, magnetite crystals with the similar shapes have been widely found in biogenic materials,[17, 18] where they form a chain with their {111} facets aligned along the length of the chain.

The growth of nanoparticles is determined by the competition between bulk energy (favors growth) and surface/interface energy (favors shrinkage).[19] The surface energy density of different crystallographic planes is usually different, and a general sequence may hold, σ{111} < σ{001} < σ{101}.[20-22] The high-index crystallography planes usually have higher surface energy. Therefore, the particles tend to be surrounded by the low-index planes and form a polyhedron shape. In fact, regular octahedra are surrounded completely by {111} planes. Taking the surface area into account, the Wagner-Seitz crystals [Fig.3 (a)] may be the reasonable shape of the magnetite nanoparticles.

As mentioned above, the magnetite nanoparticles form a bcc superlattice. Since the orientation relationships between Fe3O4 lattice and nanoparticle superlattice are also obtained, the bcc unit of the superlattice can be constructed, as shown in Fig. 3(b). All the octahedral nanoparticles in the bcc unit have the same orientation. The superlattice parameter is about 16.5 nm, which can be obtained from plane spacings of the superlattice in Fig. 1(a-c) or reciprocal vectors of the FFT patterns in Fig. 1(d-f). In fact, bcc stack of Wagner-Seitz crystals is the most close-packed [Fig. 3(c)]. Although the magnetite nanoparticles are not perfect Wagner-Seitz crystals in shape, the bcc stack should still be the most close-packed because of the polymer surface layer.

In the assembling of nanoparticles dispersed in solution, they are driven close up by the attractive force and, finally, form a stable superstructure under the balance of steric interaction between them. The attractive interaction is usually provided by van der Waals force.[8] In magnetic nanoparticles, magnetostatic interaction, which is usually stronger than van der Waals force, may exist and dominate the assembling process.[23] However, occurrence of magnetostatic interaction depends on the thermal stability of the magnetic moment of nanoparticles. If the magnetic moments cannot keep their direction during assembling, alternating electromagnetic interaction, instead of magnetostatic interaction, will occur here. Therefore, the thermal stability of magnetic moments should be studied to certify the existence of magnetostatic interaction. The superparamagnetic blocking temperature of the nanoparticle system, TB, is about 91 K measured by Quantum design MPMS-5s. According to Arrhenius law, the magnetic moments alternate their directions at the frequency of ~ 106 Hz at room-temperature. Therefore, magnetostatic interaction does not play an important role in the assembly process at room temperature.

In the absence of magnetostatic interaction, the van der Waals force solely leads the nanoparticles to aggregation. It is the repulsive steric force that stops the aggregation. Therefore, the shape of nanoparticles is an important factor by the steric force in determining the pattern of assembly. The steric force favors face-face contact, rather than edge-edge or apex-apex contact, to form a stable superstructure.[22] As well know, the size distribution of nanoparticles is critical to get a regular assembly. At the same time, shape is the dominant factor in determining the structure of a superlattice.

To self-assemble to regular superstructures is critical for the applications of nanoparticles, such as in high density magnetic data storage. As a matter of fact, self-orientation is also very important for the applications. The self-assembled and –orientated superlattices will definitely help nanoparticles advance into more practical applications. Since all the octahedral nanoparticles in the superlattice have the same orientation, the nanoparticles within the superlattice are just like divided from a whole single crystal. The self-assembled superlattice of the small particles at nanoscale is also important, as a complement to the conventional lithography.

## Conclusion

In summary, the synthesized magnetite nanoparticles have the shape of Wagner-Seitz crystals. The nanoparticles assemble to form bcc structured superlattice by themselves. The nanoparticles have the same orientation in the superlattice. The self-orientation of the nanoparticles in the self-assembled superlattice is critical to many applications of magnetic nanoparticles, such as in information storage. As a complement to conventional lithography, the self-assembly and self-orientation has wide application in various fields.

## Acknowledgment

# Reference

1. Alivisatos, A. P. et al. Organization of 'nanocrystal molecules' using DNA. Nature 382, 609-611 (1996).

2. Sun, S. H., Murray, C. B., Weller, D., Folks, L. & Moser, A. Monodisperse FePt nanoparticles and ferromagnetic FePt nanocrystal superlattices. Science 287, 1989-1992 (2000).

3. Ohara, P. C., Leff, D. V., Heath, J. R. & Gelbart, W. M. Crystallization of Opals from Polydisperse Nanoparticles. Physical Review Letters 75, 3466-3469 (1995).

4. Lalatonne, Y., Richardi, J. & Pileni, M. P. Van der Waals versus dipolar forces controlling mesoscopic organizations of magnetic nanocrystals. Nature Materials 3, 121-125 (2004).

5. Cheng, J. Y., Mayes, A. M. & Ross, C. A. Nanostructure engineering by templated self-assembly of block copolymers. Nature Materials 3, 823-828 (2004).

6. Yamamuro, S., Farrell, D. F. & Majetich, S. A. Direct imaging of self-assembled magnetic nanoparticle arrays: Phase stability and magnetic effects on morphology. Physical Review B 65, - (2002).

7. Farrell, D. et al. Dipolar interactions and structural coherence in iron nanoparticle arrays. Journal of Magnetism and Magnetic Materials 282, 1-5 (2004).

8. Narayanan, S., Wang, J. & Lin, X. M. Dynamical self-assembly of nanocrystal superlattices during colloidal droplet evaporation by in situ small angle x-ray scattering. Physical Review Letters 93, - (2004).

9. Israelachvili, J. N. Intermolecular and surface forces (Academic Press, London ; San Diego, 1991).

10. *Dumestre, F., Chaudret, B., Amiens, C., Renaud, P. & Fejes, P. Superlattices of iron nanocubes synthesized from Fe[N(SiMe3)(2)](2). Science 303, 821-823 (2004).*

11. *Shenton, W., Pum, D., Sleytr, U. B. & Mann, S. Synthesis of cadmium sulphide superlattices using self-assembled bacterial S-layers. Nature 389, 585-587 (1997).*

12. *Redl, F. X., Cho, K. S., Murray, C. B. & O'Brien, S. Three-dimensional binary superlattices of magnetic nanocrystals and semiconductor quantum dots. Nature 423, 968-971 (2003).*

13. *Murray, C. B., Kagan, C. R. & Bawendi, M. G. Self-Organization of Cdse Nanocrystallites into 3-Dimensional Quantum-Dot Superlattices. Science 270, 1335-1338 (1995).*

14. *Sun, S. H. & Zeng, H. Size-controlled synthesis of magnetite nanoparticies. Journal of the American Chemical Society 124, 8204-8205 (2002).*

15. *Sun, S. H. et al. Monodisperse MFe2O4 (M = Fe, Co, Mn) nanoparticles. Journal of the American Chemical Society 126, 273-279 (2004).*

16. *Pileni, M. P. Nanocrystal self-assemblies: Fabrication and collective properties. Journal of Physical Chemistry B 105, 3358-3371 (2001).*

17. *Kirschvink, J. L., Kobayashikirschvink, A. & Woodford, B. J. Magnetite Biomineralization in the Human Brain. Proceedings of the National Academy of Sciences of the United States of America 89, 7683-7687 (1992).*

18. *Thomas-Keprta, K. L. et al. Truncated hexa-octahedral magnetite crystals in ALH84001: Presumptive biosignatures. Proceedings of the National Academy of Sciences of the United States of America 98, 2164-2169 (2001).*

19. *Zhang, K.-Q. & Liu, X. Y. Nature 429, 739 (2004).*

20. Wang, Z. L. Polyhedral shapes of CeO2 nanoparticies. *Journal of Physical Chemistry B* **104**, 1353 (2000).

21. Renaud, L. P., Blume, H. W., Pittman, Q. J., Lamour, Y. & Tan, A. T. Thyrotropin-Releasing-Hormone Selectively Depresses Glutamate Excitation of Cerebral Cortical-Neurons. *Science* **205**, 1275-1277 (1979).

22. Wang, Z. L. Transmission Electron Microscopy of Shape-Controlled Nanocrystals and Their Assemblies. *J. Phys. Chem. B* **104**, 1153-1175 (2000).

23. Pileni, M. P. The role of soft colloidal templates in controlling the size and shape of inorganic nanocrystals. *Nature Materials* **2**, 145-150 (2003).

**Figure captions**

Fig. 1

(a-c) are the [001], [101], and [111] projections of the magnetite nanoparticle superlattices, respectively; (d-f) are the corresponding FFT patterns of the superlattice shown in (a-c); (g-i) are the corresponding electron diffraction of $Fe_3O_4$ shown in (a-c).

Fig. 2

(a-c) are the HRTEM images of [001], [101], and [111] oriented truncated octahedral nanoparticles, respectively; (d-f) are the corresponding FFT patterns of the HRTEM images shown in (a-c); (g-i) show the shapes of [001], [101], and [111] oriented Wagner-Seitz crystals.

Fig. 3

(a) Wagner-Seitz crystals; (b) The schematic diagram of a bcc unit of the superlattice composed of Wagner-Seitz crystalss.

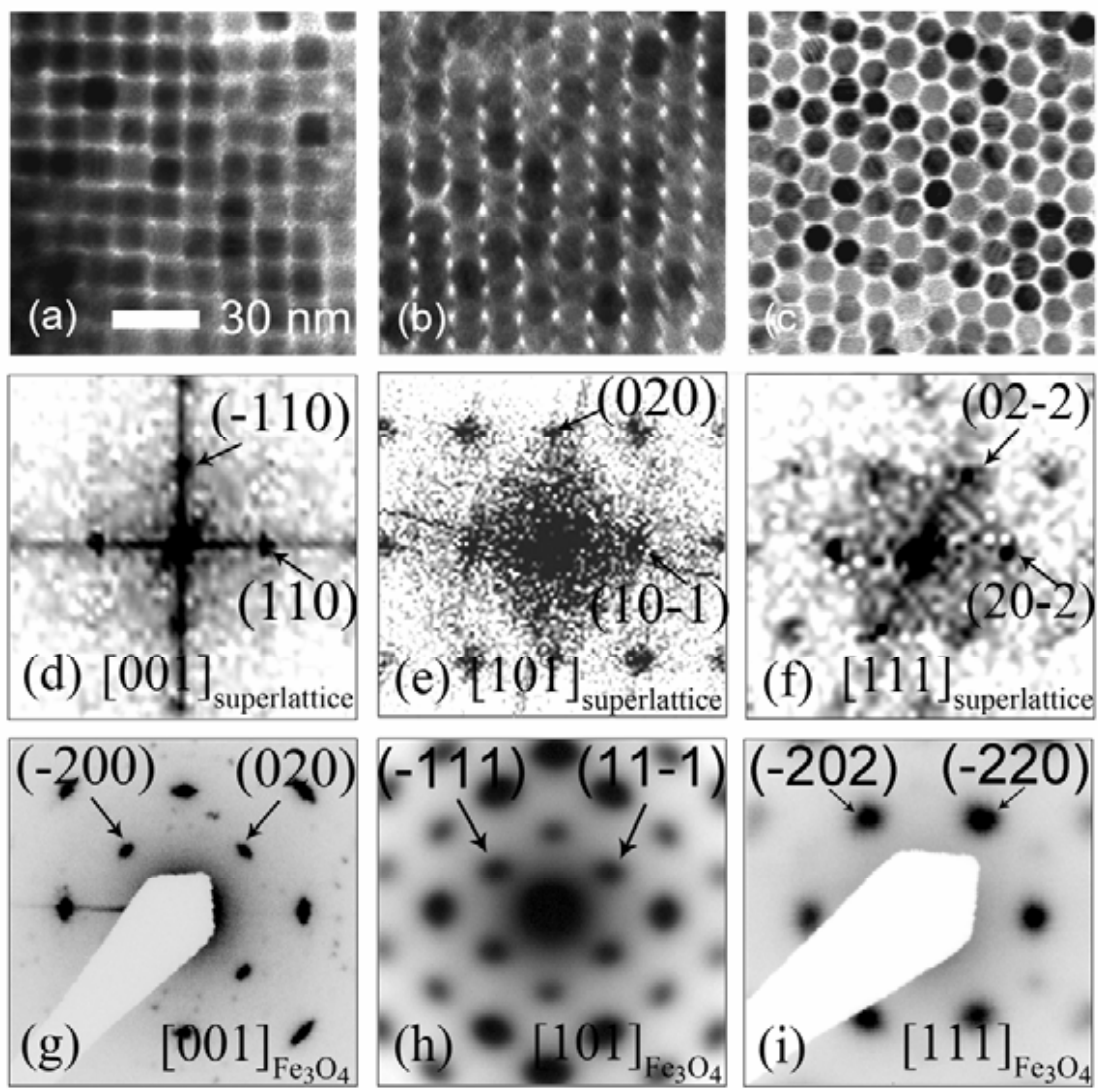

Fig. 1

Zheng et al

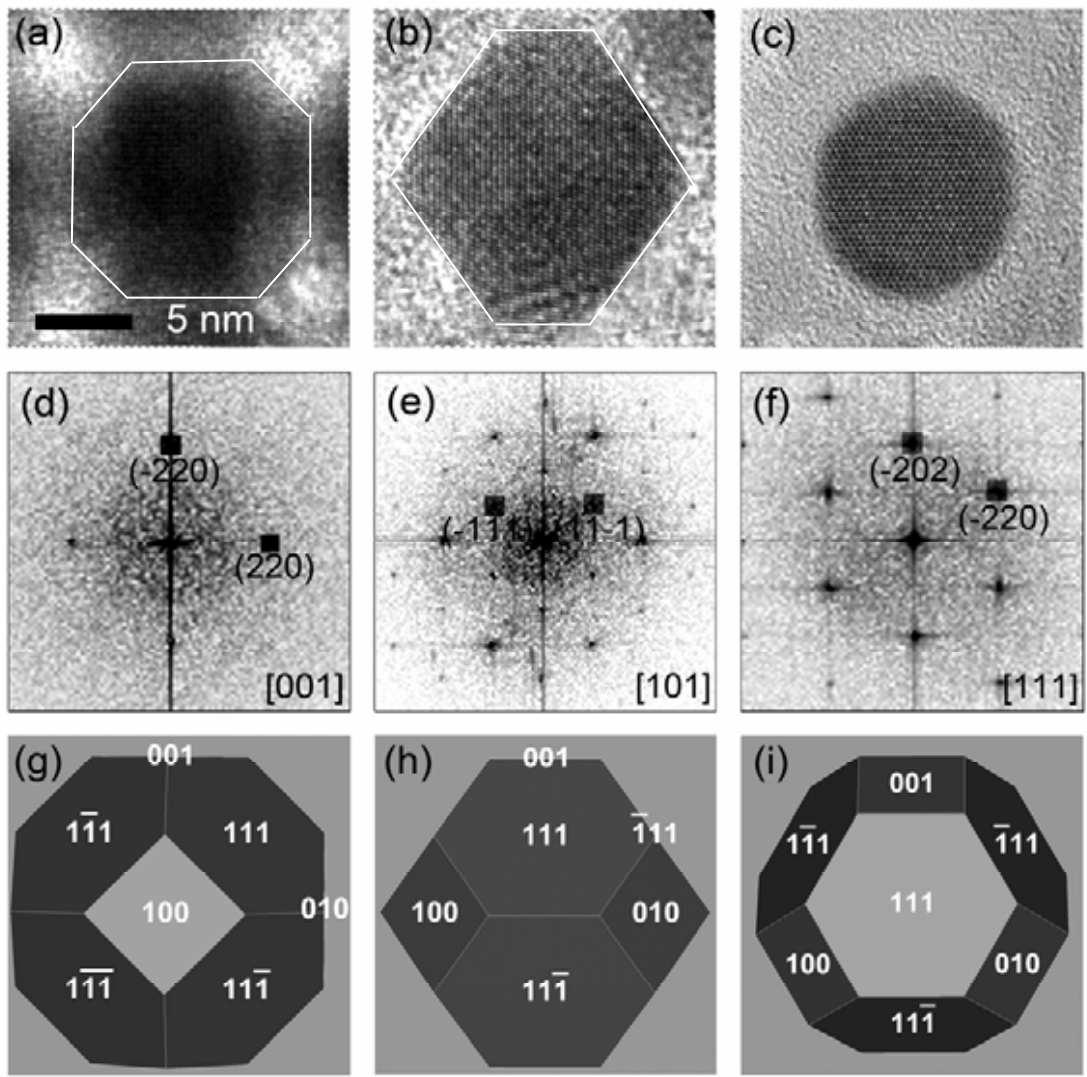

Fig. 2

Zheng et al

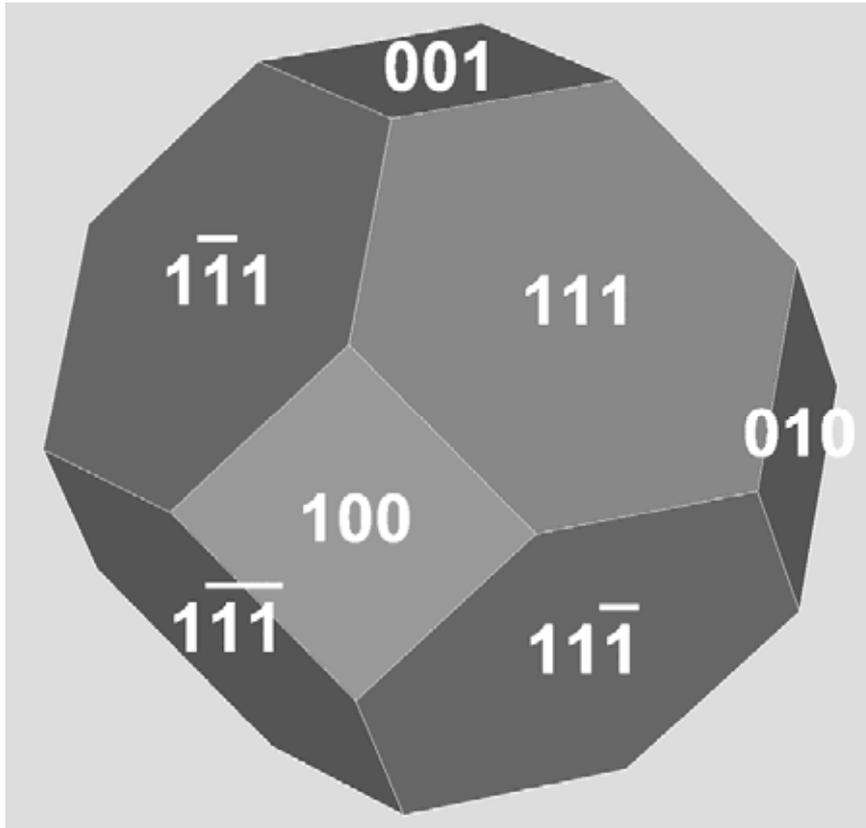

*Fig. 3(a)*

*Zheng et al*

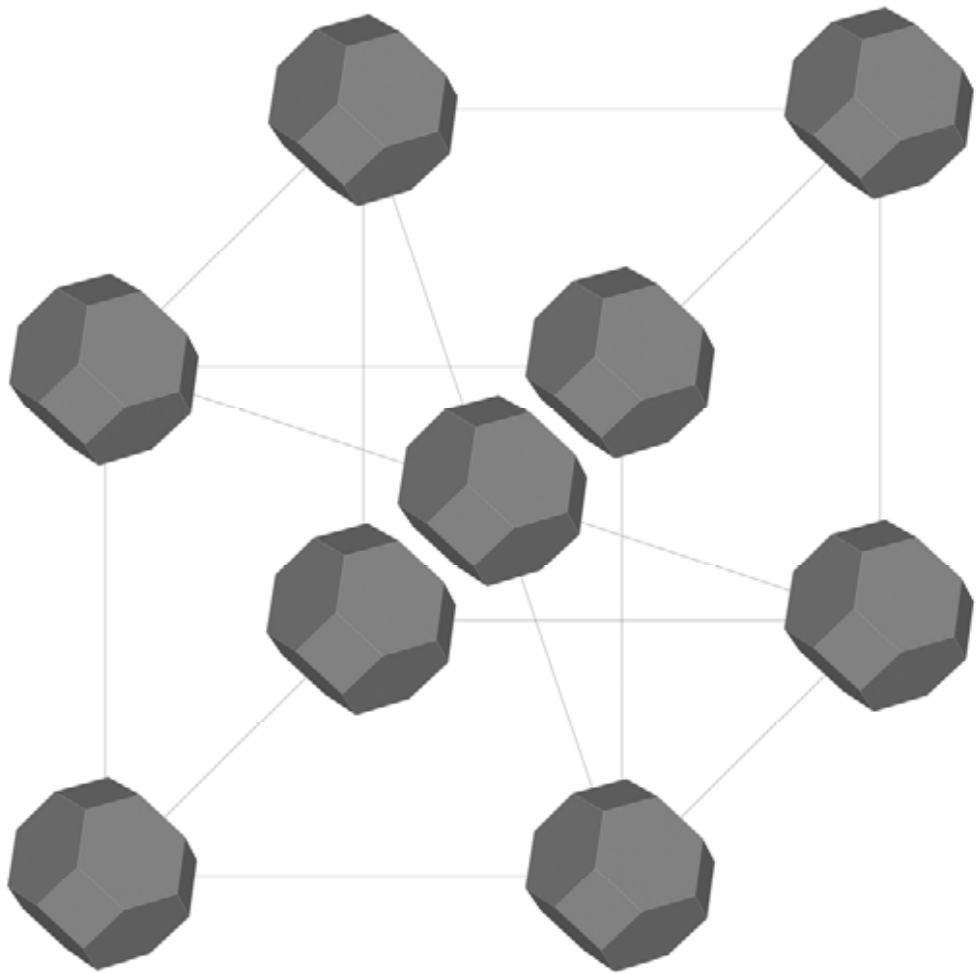

*Fig. 3(b)*

*Zheng et al*